\documentclass[conference]{IEEEtran}
\IEEEoverridecommandlockouts
\usepackage{cite}
\usepackage{amsmath,amssymb,amsfonts}
\usepackage{algorithm,algorithmic}
\usepackage{graphicx}
\usepackage{textcomp}
\usepackage{xcolor}
\usepackage{tikz}
\usetikzlibrary {arrows.meta}
\usepackage{pgfplots}
\pgfplotsset{compat=newest}

\usepackage{fancyhdr}
\makeatletter
\def\ps@IEEEtitlepagestyle{
    \def\@oddfoot{\strut\hfill\parbox{\textwidth}{This work has been submitted to the IEEE for possible publication. Copyright may be transferred without notice, after which this version may no longer be accessible.}\hfill}%
    \def\@evenfoot{}%
    \def\@oddhead{}%
    \def\@evenhead{}%
}
\makeatother

\usepackage{url}

\usepackage{threeparttable}

\usepackage{subfigure}

\usepackage{multirow}

\newcommand{\blank}{\vspace{-0.3cm}}

\def\BibTeX{{\rm B\kern-.05em{\sc i\kern-.025em b}\kern-.08em
    T\kern-.1667em\lower.7ex\hbox{E}\kern-.125emX}}
\begin{document}

\title{L-Sort: An Efficient Hardware for Real-time Multi-channel Spike Sorting with Localization
\thanks{This work was supported in part by the Royal Society under grant IEC\textbackslash NSFC\textbackslash 223067.}
}

\author{\IEEEauthorblockN{Yuntao Han, Shiwei Wang, Alister Hamilton}
\IEEEauthorblockA{
Institute for Integrated Micro and Nano Systems, School of Engineering, University of Edinburgh, Edinburgh, UK\\
\{s2328236, shiwei.wang, Alister.Hamilton\}@ed.ac.uk}
}

\maketitle

\begin{abstract}
Spike sorting is essential for extracting neuronal information from neural signals and understanding brain function. With the advent of high-density microelectrode arrays (HDMEAs), the challenges and opportunities in multi-channel spike sorting have intensified. Real-time spike sorting is particularly crucial for closed-loop brain computer interface (BCI) applications, demanding efficient hardware implementations. This paper introduces L-Sort, an hardware design for real-time multi-channel spike sorting. Leveraging spike localization techniques, L-Sort achieves efficient spike detection and clustering without the need to store raw signals during detection. By incorporating median thresholding and geometric features, L-Sort demonstrates promising results in terms of accuracy and hardware efficiency. We assessed the detection and clustering accuracy of our design with publicly available datasets recorded using high-density neural probes (Neuropixel). We implemented our design on an FPGA and compared the results with state of the art. Results show that our designs consume less hardware resource comparing with other FPGA-based spike sorting hardware. 
\end{abstract}


\section{Introduction}

Spike sorting is an algorithm used to distinguish and classify the action potentials of individual neurons from recorded neural signals, enabling the detailed analysis of neural activity and the improvement of brain-computer interface (BCI) performance by accurately interpreting neuronal information~\cite{ah_survey}. Conventionally, this algorithm follows the pipeline of spike detection, feature extraction, and clustering. 
Several studies explored the algorithm adaption and tailored hardware for multi-channel spike sorting, as shown in TABLE~\ref{table:ss}. 

Currently, high-density microelectrode arrays (HDMEA) and neural probes, e.g., Neuropixel 2.0~\cite{neuropixel_2} and BioCam4096~\cite{biocam_ncl}, are capable of recording with over thousands of channels with the electrode pitch shrunk down to tens of micrometers. Therefore, the activity of most neurons can be detected across multiple channels. This multiplicity increases the potential for erroneous identification of redundant spikes but also provides additional geometric information about the spatial distribution of neuronal activity. Utilising these geometric information, spike localization~\cite{localization_ori} is a recently emerging method which infers the physical position of the spike source from multi-channel recordings with electrode geometry. The calculated geometric information is then utilized for succeeding clustering. As the spikes are captured by multiple electrodes, they can be localised based on the relative voltage amplitudes of these channels. A widely used technique for spike localization is the center of mass (CoM) method~\cite{localization_ori}, which has been demonstrated capable of distinguishing different spike sources~\cite{localization_cell}. This method computes the source position by taking the weighted average of the positions of a selected set of channels. The weights are the amplitudes of these channels, typically including all surrounding channels centered around the one with the highest amplitude. Comparing with the calculation of conventional features, e.g., first-and-second-derivative-extrema (FSDE)~\cite{geo_osort_fpga}\cite{geo_osort_asic} which involves the whole recording in the time window, these positions can be calculated with very few timesteps, eliminating the need for retaining all the recorded waveforms in the time window.

\begingroup
\renewcommand{\arraystretch}{1.2}
\begin{table}[t]
\centering
\caption{Recent Works of Multi-channel Spike Sorting}
\label{table:ss}
\resizebox{\linewidth}{!}{%
\begin{threeparttable}
\setlength\tabcolsep{3pt}
\begin{tabular}{lllll}
\hline \hline
\textbf{Work}                        & \textbf{Platform} & \textbf{Detection} & \textbf{Feature}     & \textbf{Clustering} \\ \hline
NIPS'2019~\cite{localization_nips19} & GPU               & -                  & 2D-geometry          & AVI                 \\
NIPS'2021~\cite{localization_nips21} & GPU               & -                  & 3D-geometry$^1$      & point-cloud         \\ \hline
TBCAS'2019~\cite{posort}     & FPGA/ASIC         & mean-TH            & -                    & O-Sort              \\
Access'2020~\cite{zyon}    & FPGA              & amplitude-TH       & FSDE                 & K-Means             \\
TBME'2020~\cite{geo_osort_fpga}      & FPGA              & mean-TH            & -                    & O-Sort$^2$          \\
TBCAS'2023~\cite{detection_ic}       & FPGA/ASIC         & median-TH          & -                    & -                   \\
JSSC'2023~\cite{geo_osort_asic}      & ASIC              & mean-TH            & peak-FSDE            & O-Sort$^2$          \\ \hline
\textbf{This work}                        & \textbf{FPGA}     & \textbf{median-TH} & \textbf{2D-geometry} & \textbf{O-Sort}     \\ \hline \hline
\end{tabular}%
\begin{tablenotes}
    \item[1] with an additional dimension calculated with triangulation.
    \item[2] explored locality with geometry information.
\end{tablenotes}
\end{threeparttable}
}
\blank
\end{table}
\endgroup

In this paper, we propose L-Sort, an efficient hardware design targeting real-time multi-channel spike sorting with median-based peak detection, geometric feature extraction (spike localization), and O-Sort clustering. To our best knowledge, this work presents the first localization-based spike sorting accelerator implemented on a miniaturized chip hardware (FPGA). Our contributions are summarized as follows:

\noindent1) We designed an efficient hardware architecture for peak\footnote{Since each spike generated by neurons could be captured by several channels, we refer each detected outlier in each channel as \emph{peak} instead of spike for clearance.} detection using median-based threshold. To reduce the hardware consumption for finding medians, we keep the sorted sequence from last timestep and incrementally insert the new sample into this sequence to reduce the complexity.
 
\noindent 2) We proposed a peak-based CoM algorithm to localize the spike source with the detected peaks instead of all surrounding channels, eliminating the requirement on keeping whole recording during spike sorting. The corresponding hardware uses buffers to group detected peaks into spikes.
  
\noindent3) We assessed the performance of L-Sort through implementing on FPGA board. Experiment results show that our design could significantly reduce hardware consumption comparing with other hardware designs for real-time multi-channel spike sorting. Detailed analyses on median calculators and memory consumption are also presented for demonstrating the improvements provided by our proposed hardware design.

\section{Methods}
\label{sec:method}

\subsection{Overview}
\begin{figure}[t]
  \centering
  \includegraphics[width=1.0\linewidth]{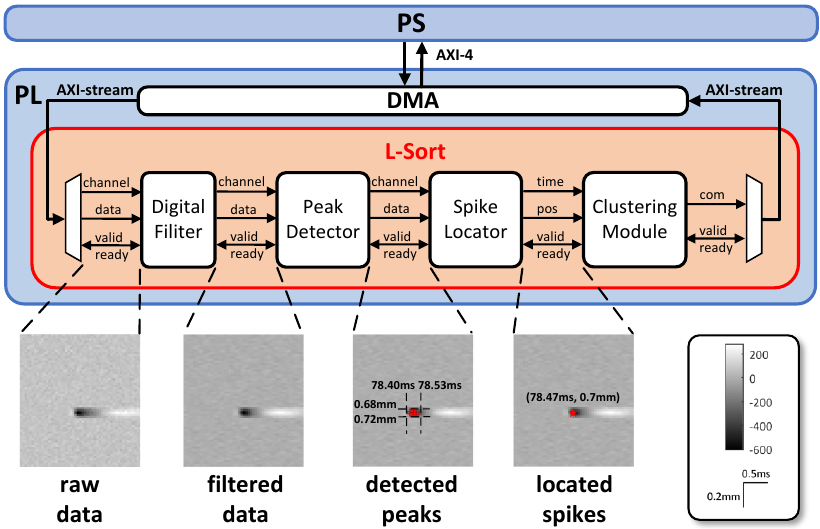}
  \caption{Overview of proposed hardware with example recording sampled from the dataset ('Set 1' in \cite{dataset_neuropixel}). This recording encompasses all $60$ channels where $x=0$. The time interval spans from $76.67ms$ to $80.00ms$, during which a spike occurs at $78.50ms$ according to the provided ground truth.}
  \label{fig:overview}
  \blank
\end{figure}

The hardware design of L-Sort is constructed with digital filter, peak detector, spike locator, and clustering module, as shown in Fig.~\ref{fig:overview}. These modules are interfaced using AXI-Stream protocol. We also build a closed-loop system for sending raw recordings and receiving results of spike sorting with the embedded CPU, a.k.a., processing system (PS), interacting with the designed system through AXI-DMA.

The process begins with PS sending raw recordings to L-Sort hardware with channel indices and voltage values. The digital filter, as the first module of L-Sort, removes the local field potentials and high-frequency noises from the raw data. The filtered data is then processed by the peak detector to find the peaks in different channels based on median thresholding. These peaks are grouped into spikes and the locations of these spikes are calculated as geometric features. Finally, the clustering module utilizes these positions to classify these spikes to different neurons with O-Sort algorithm, and sends the sorted spikes to PS with timestamps and neuron indices.

\subsection{Digital Filter}
The digital filter is a first-order infinite-impulse-response (IIR) filter with band-pass frequency from $300$ to $6000$ Hz. The implementation of this filter follows the Direct Form \uppercase\expandafter{\romannumeral2}~\cite{wp330}, whose coefficients are quantized to $12$-bit signed and fixed representation with $10$ bits for fractions.

\subsection{Peak Detector}
The peak detector finds samples with relatively high absolute values from the filtered data through channel-wise median thresholding. The threshold $TH[c,t]$ is calculated as follows:
\begin{equation}
\begin{aligned}
    TH[c,t]&=M\times median(x[c,t-N:t])
\end{aligned}
\end{equation}
where $c$ and $t$ are channel indices and timestamps, $N$ is the number of points used for calculating median. Median-abased thresholding is more robust to outliers compared with using means~\cite{detection_ic}. However, finding medians is more hardware-consuming than calculating means, as the sorting of sequence requires comparing each point with each point, resulting in an $O(n^2)$ algorithm complexity. Media recursion methods have been explored to improve hardware efficiency \cite{detection_ic}, which estimats the real median with median-of-median to reduce the number of points involved in each median calculation.

To facilitate streamed recordings, we propose to calculate the median in an incremental manner, i.e., the points considered for the new median are composed of the new sample and the points used in the last median except the oldest point. It is possible to utilize the previously sorted sequence, along with removing the oldest point and inserting the new sample, to sort the new points. This method could simplify the median finding from comparing each point with each point to comparing only the new point with each point, i.e., from $O(n^2)$ to $O(n)$.
\begin{figure}[t]
  \centering
  \includegraphics[width=1.0\linewidth]{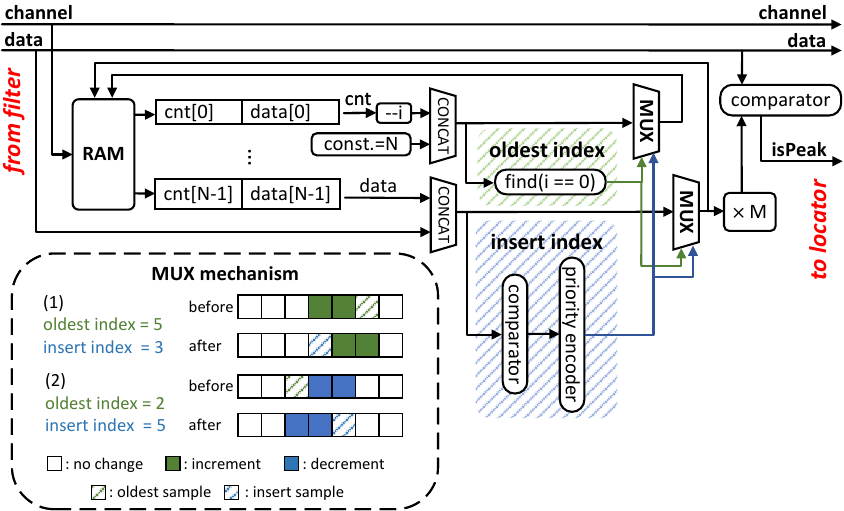}
  \caption{Hardware architecture of peak detection based on median thresholding. $N$ represents number of points used for calculating median, and $M$ represents the factor for calculating threshold.}
  \label{fig:median}
  \blank
\end{figure}

The hardware implementation of the proposed incremental median finding is shown in Fig.~\ref{fig:median}. Our proposed incremental method requires finding the oldest index and insert index, which are the indices for the oldest sample and the new sample, respectively. In our implementation, the previous points are stored in a RAM including their amplitude values and associated counters for tracking their temporal order. The counter of the new sample is set to $N$ and decreased by $1$ after each RAM access. The oldest index is acquired when the value of the counter becomes $0$ which indicates that the corresponding sample is the oldest one of the stored points. The insert index is calculated by comparing the new sample with all previous points. These two indices are used for controlling multiplexers to sort the current points, which are written back to the RAM and indexed to find the median. This median is used to calculate the threshold, which is compared with the new sample to identify the presence of a peak.
\subsection{Spike Locator}
The spike locator is responsible for grouping the detected peaks into spikes and localizing these spikes. Unlike existing CoM-based localization techniques considering all surrounding channels to calculate the position of spike source, our proposed peak-based CoM method utilizes only the detected peaks, thus eliminating the need to access amplitudes of all surrounding channels. Assuming that the probe is placed in an XZ plane and the detected peaks are $P$, the peak-based CoM calculates the position of spike $[X_{spike}, Z_{spike}]$ as follows:
\begin{equation}
\begin{aligned}
    X_{spike} = \frac{\sum_{p\in P}amp_{p}x_{p}}{\sum_{p\in P}amp_{p}} \\
    Z_{spike} = \frac{\sum_{p\in P}amp_{p}z_{p}}{\sum_{p\in P}amp_{p}}
\end{aligned}
\end{equation}

\begin{figure}[t]
  \centering
  \includegraphics[width=1.0\linewidth]{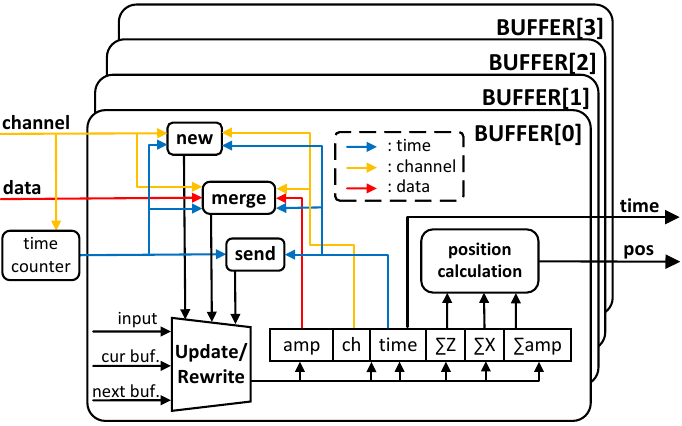}
  \caption{Hardware architecture of spike locator based on modified CoM with buffer size of $4$, where $\sum Z$, $\sum X$, and $\sum amp$ represents $\sum amp_{p}z_{p}$, $\sum amp_{p}x_{p}$, and $\sum amp_{p}$, respectively.}
  \label{fig:locator}
  \blank
\end{figure}

\begin{figure}[t]
  \centering
  \includegraphics[width=0.95\linewidth]{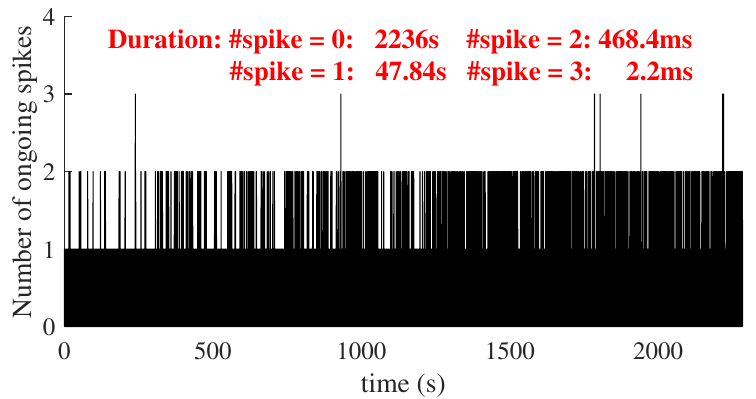}
  \caption{Ongoing spike count in each timestep (data from Set 1 in \cite{dataset_neuropixel}).}
  \label{fig:spike}
  \blank
\end{figure}

The hardware architecture of the spike locator is shown as Fig.~\ref{fig:locator}. Because of the temporal overlaps among different spikes for multi-channel recordings, we implement a buffer with size of $4$ to store up to $4$ spikes based on our observation of up to $3$ ongoing spikes in each time step as shown in Fig.~\ref{fig:spike}. Each buffer can record information of an ongoing spike including the time, channel, and amplitude of the peak with the highest amplitude as well as the sums for localization. These contents can be changed by the coming peaks. When a peak is fed into the spike locator, each buffer will check if the differences between the input spike with the recorded peak channel and time are within pre-defined thresholds, and determine whether this peak should be merged into an existing spike or identified as a new spike. If it is the former case, the corresponding sums are increased and the peak information is updated if the new peak has a higher amplitude than the recorded ones. Otherwise, a new spike identification is generated and assigned with a new buffer. When there is no peak received, the first buffer will check the time elapsed after the newest merged peak. If the time interval exceeds the threshold for merging, which indicates that all the peaks of the recorded spike have been received, the first buffer will calculate the position and send it together with the recorded peak time. Meanwhile, each buffer except the last one will update its contents with the contents from the next buffer.

\subsection{Clustering Module}
The clustering of the detected spikes are performed with O-Sort scheme~\cite{posort}, which mainly includes 1) merging spike into an existing cluster or creating a new cluster and 2) updating the cluster and merging it to other clusters if the updated cluster is similar to another cluster. For conventional O-Sort coupling with FSDE features, the threshold for determining the merging of spike or cluster to cluster is calculated on the fly. By contrast, the localization-based features have a realistic meaning in biology, i.e., the positions of spike sources. The intervals among these sources can be estimated based on the monitored neural tissue area. Therefore, we use a fixed merging threshold to reduce the hardware resource utilization.

\section{Results}

The proposed L-Sort hardware was implemented on a ZCU102 board, which is equipped with a Zynq UltraScale+ XCZU9EG MPSoC containing both FPGA and a quad-core ARM Cortex-A53 processor for PL and PS sides, respectively. The PS side is configured with the PetaLinux provided by Xilinx and communicated with the host PC through Ethernet for PL configuration and result visualization. The raw recording is saved in the SD card and read into DDR by PS before transmitted to the PL side for spike sorting.
 \begin{figure}[t]
   \centering
   \includegraphics[width=0.95\linewidth]{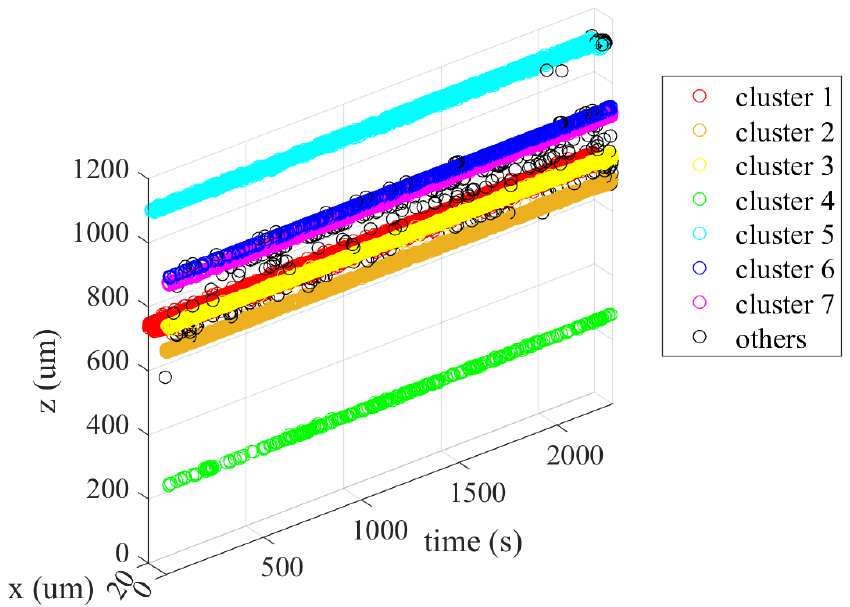}
   \caption{Spike clustering of Set 1 in \cite{dataset_neuropixel}. Clusters matched with the seven clusters provided in the ground truth are marked with non-black colors, while others are marked with black.}
   \label{fig:cluster}
   \blank
 \end{figure}
\subsection{Software Validation of Algorithm Accuracy}
We use the real data~\cite{dataset_neuropixel} recorded by Neuropixel to assess the accuracy of the proposed spike sorting algorithm. This dataset contains $120$ channels and $7$ identified neurons. The results are shown in Fig.~\ref{fig:cluster}. Our algorithm achieves $97.19\%$ and $96.93\%$ detection and classification accuracy, respectively.

\subsection{Hardware Utilization}
The hardware utilization of the L-Sort is shown in TABLE~\ref{table:com_hw}. It shows that our implementation consumes significantly less hardware resource. While others require complex calculation to determine the threshold for clustering and require keeping the spike train during spike sorting, our implementation with the modified spike localization can empirically set the clustering threshold based on the geometric interval of neurons and extract the positions of spike sources with the detected peaks instead of accessing the whole recording.
\begingroup
\renewcommand{\arraystretch}{1.1}
\begin{table}[]
\centering
\caption{Hardware Utilization of L-Sort}
\label{table:com_hw}
\resizebox{\linewidth}{!}{%
\setlength\tabcolsep{2pt}
\begin{tabular}{llllllllll}
\hline \hline
                               & \textbf{}      & \multicolumn{3}{l}{\textbf{Input Dataset}}                                                                        & \multicolumn{5}{l}{\textbf{FPGA Utilization}}                                                                                             \\ \cline{3-10} 
                               &                & \textbf{Bitwidth} & \textbf{\begin{tabular}[c]{@{}l@{}}Sampling\\ Rate\\ (kHz)\end{tabular}} & \textbf{\#channel} & \textbf{LUT}    & \textbf{FF}     & \textbf{BRAM} & \textbf{DSP}  & \textbf{\begin{tabular}[c]{@{}l@{}}Clock\\ Rate\\ (MHz)\end{tabular}} \\ \hline
\multicolumn{2}{l}{TBCAS'2019~\cite{posort}}                  & -                & -                          & -                                                          & $16472$         & $8444$          & $29$          & $130$         & $123$                                                                 \\
\multicolumn{2}{l}{Access'2020~\cite{zyon}}                 & $12$              & $18$                                                                     & $4096$             & $26444$         & $28944$         & $104$         & $61$          & $125$                                                                 \\
\multicolumn{2}{l}{TBME'2020~\cite{geo_osort_fpga}}                   & $16$              & $20$                                                                     & $128$              & $17484$         & $51674$         & $98$          & $60$          & $200$                                                                 \\ \hline
\multirow{5}{*}{\textbf{Ours}} & \multicolumn{4}{l}{Digital Filter}                                                                                                 & $51$            & $46$            & $0.5$         & $4$           &                                                                       \\
                               & \multicolumn{4}{l}{Peak Detector}                                                                                                  & $1237$          & $84$            & $5.5$         & $0$           &                                                                       \\
                               & \multicolumn{4}{l}{Spike Locator}                                                                                                  & $1933$          & $590$           & $0$           & $9$           &                                                                       \\
                               & \multicolumn{4}{l}{Clustering Module}                                                                                              & $3387$          & $2297$          & $0$           & $2$           &                                                                       \\
                               & \textbf{total} & $12$              & $30$                                                                     & $120$              & $\mathbf{6513}$ & $\mathbf{3017}$ & $\mathbf{6}$  & $\mathbf{15}$ & $3.6$                                                                 \\ \hline \hline
\end{tabular}
}
\blank
\end{table}
\endgroup

\subsubsection{Median-based Detection}
\begin{figure}
\centering 
\subfigure[for logic]
{%
\begin{tikzpicture}
\begin{axis}[
    font=\scriptsize,
    width=0.5\linewidth,
    height=0.5\linewidth,
    xlabel={\#Points},
    ylabel={Number of LUTs},
    xtick={25,50},
    grid=major,
]


\addplot[olive,mark=*,mark size=3pt,line width=2pt] coordinates {
    (25, 1237)
    (50, 2312)
}; \label{line:median_unroll}

\addplot[teal,mark=*,mark size=3pt,line width=2pt] coordinates {
    (25, 468)
    (50, 468)
}; \label{line:median_roll}

\addplot[violet,mark=triangle*,mark size=3pt,line width=2pt] coordinates {
    (25, 7500)
}; \label{line:median_real}

\addplot[red,mark=triangle*,mark size=3pt,line width=2pt] coordinates {
    (25, 1278)
    (50, 2394)
}; \label{line:median_hw_this work}

\end{axis}
\end{tikzpicture}
}
\subfigure[for memory]
{%
\begin{tikzpicture}
\begin{axis}[
    font=\scriptsize,
    width=0.5\linewidth,
    height=0.5\linewidth,
    xlabel={\#Points},
    ylabel={Bitwidth of RAMs},
    xtick={25,50},
    grid=major,
]

\addplot[olive,mark=*,mark size=3pt,line width=2pt] coordinates {
    (25, 500)
    (50, 1000)
}; \label{line:median_unroll}

\addplot[teal,mark=*,mark size=3pt,line width=2pt] coordinates {
    (25, 500)
    (50, 1000)
}; \label{line:median_roll}

\addplot[violet,mark=triangle*,mark size=3pt,line width=2pt] coordinates {
    (25, 0)
}; \label{line:median_real}

\addplot[red,mark=triangle*,mark size=3pt,line width=2pt] coordinates {
    (25, 525)
    (50, 1050)
}; \label{line:median_hw_this_work}

\end{axis}
\end{tikzpicture}
}
\label{fig:median_comparison}
\caption{Hardware utilization of proposed accurate median-based detector (\ref{line:median_hw_this_work}) comparing with real (\ref{line:median_real}) and approximate (roll:\ref{line:median_roll}, unroll:\ref{line:median_unroll}) median calculators reported in~\cite{detection_ic}.} 
\blank
\end{figure}
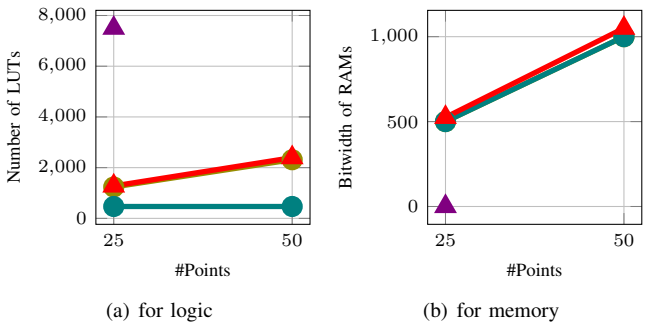
To demonstrate the hardware efficiency of the proposed hardware for calculating median, we compare our design with the results achieved from~\cite{detection_ic} which  applied the median recursion method to approximately estimate the real median by calculating median of medians to reduce the complexity, and achieved a significant reduction in LUT utilization. Our proposed median-based detector utilizes the sorted sequence from the last timestep to further reduce the complexity without approximation on the algorithm, and therefore could calculate the real median at the similar cost of the approximate median calculator.
\subsubsection{Memory Utilization and Access}
\begingroup
\renewcommand{\arraystretch}{1.1}
\begin{table}[t]
\centering
\caption{Memory Utilization and Access}
\label{table:memory}
\begin{tabular}{lll}
\hline \hline
Work               & Utilization (KB)         & Access (bit/spike)     \\ \hline
JSSC'2023~\cite{geo_osort_asic}          & 97.96                    & 6696                   \\
\textbf{This work} & \textbf{25.52 (26.2\%)} & \textbf{1140 (17.0\%)} \\ \hline \hline
\end{tabular}
\blank
\end{table}
\endgroup
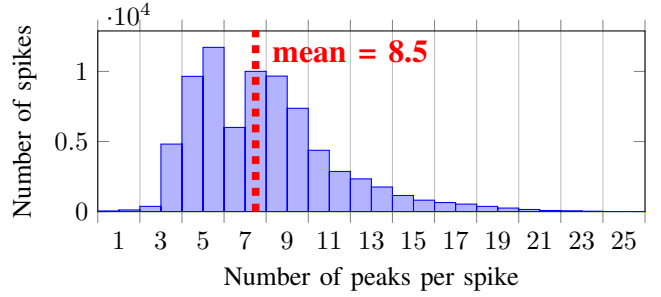
\begin{figure}[t]
\centering
\begin{tikzpicture}
\begin{axis}[width=\linewidth,
             height=0.45\linewidth,
             ybar interval,
             ymin=0,
             xmin=1,
             xmax=27,
             xtick={1,3,5,...,27},
             xlabel=Number of peaks per spike,
             ylabel=Number of spikes]
\addplot coordinates {
(1, 49)
(2, 121)
(3, 377)
(4, 4817)
(5, 9647)
(6, 11714)
(7, 6005)
(8, 10004)
(9, 9666)
(10, 7370)
(11, 4375)
(12, 2866)
(13, 2336)
(14, 1756)
(15, 1152)
(16, 817)
(17, 650)
(18, 537)
(19, 374)
(20, 255)
(21, 154)
(22, 82)
(23, 60)
(24, 30)
(25, 9)
(26, 8)
(27, 9)
};
\draw[red, dashed, line width=1mm] (axis cs:8.5,0) -- (axis cs:8.5,13000);
\node[red, font=\bfseries\large] at (axis cs:13,10000) [anchor=south] {mean = 8.5};
\end{axis}
\end{tikzpicture}
\caption{
Histogram of number of detected peaks for each spike.
}
\label{fig:memory}
\blank
\end{figure}
To verify the advantage of utilizing geometric features compared with conventional method that requires access to whole recording, we quantitatively compare our design with \cite{geo_osort_asic}. For a fair comparison, we modified our design to make it capable of processing same number of channels with \cite{geo_osort_asic}, i.e., $384$ channels. The results are shown in TABLE~\ref{table:memory}. Overall, \cite{geo_osort_asic} implemented $97.96$ kB memory on chip while ours only implemented $25.52$ kB, saving $73.8\%$ of area for memory. The majority of their memory was consumed by the $76.6$ kB input buffer, which stores the recording within the time window for spike alignment and feature extraction, while our design does not need to keep these samples. The access to the input buffer is $5616$ and $1080$ bits for finding the channel with the strongest response to the spike and calculating the FSDE features. In our design, we only need to access the $134$-bit-width buffer in spike locator to accumulate the amplitudes and amplitude-weighted locations for CoM-based localization, resulting in only $1140$ bits of memory access with an $8.5$ peaks per spike on average, as shown in Fig.~\ref{fig:memory}.

\section{Conclusion}
In this paper, we presented L-Sort, the hardware implementation of real-time multi-channel spike sorting with localization technique. We propose an efficient median-based spike detection module to find the peaks from the recording, followed by a modified localization algorithm as feature extraction and O-Sort for clustering. 
The designed median calculator could calculate the real median consuming similar hardware resource of the approximate median calculator, and the localization-based spike sorting could significantly reduce the memory utilization and access, thus increasing the hardware efficiency.

\bibliographystyle{IEEEtran}
\bibliography{L-Sort}

\end{document}